# Investigation of the structure of β-Tantalum

Aiqin Jiang[1], Anto Yohannan[1*], Neme O. Nnolim[1], Trevor A. Tyson[1], Lisa Axe[2], Sabrina L. Lee[3] and Paul Cote[3]

[1]Department of Physics, New Jersey Institute of Technology, Newark, NJ 07102

[2]Department of Civil and Environmental Engineering, New Jersey Institute of Technology, Newark, NJ 07102

[3]US Army Armament Research, Development and Engineering Center, Benét Laboratories, Watervliet, NY 12189

## Abstract

The local structure of β-tantalum was investigated by comparing experimental extended x-ray absorption fine structure (EXAFS) measurements with calculated spectra of proposed models. Four possible structure candidates were examined: a β-Uranium based structure, a distorted A15 structure, a bcc-Ta based superlattice structure with N interstitials and a simple hcp structure. The local structural measurements were found to be consistent with the β-Uranium based model containing 30 atoms per unit cell and having the space group *P4$_2$/mnm*. The thermal effect analysis on x-ray diffraction and EXAFS spectra, which reveals that β-Ta is highly disordered, agrees with the low symmetry and anisotropic system of the β-U model.

# I. INTRODUCTION

In sputter deposition experiments, tantalum thin films exhibit two crystalline phases, body-centered cubic (α-Ta) and tetragonal (β-Ta) tantalum, which differ in both structural and electrical properties. The beta phase of tantalum has been attracting much interest since its discovery in 1965 by Read and Altman[1]. Because of its high resistivity (180-220 μ.ohm.cm), it is the preferred form for fabricating resistors and capacitors[2,3]. It is also a promising underlying layer to promote the adhesion of copper to dielectrics and diffusion barrier of copper on silicon[4,5].

A great deal of work has been devoted to β-Ta, but its structure has not been well characterized. Numerous crystal structures have been reported for β-Ta. Read and Altman[1] proposed a tetragonal unit cell Ta with $a$=5.34, $c$=9.94 Å, containing 16 atoms with a theoretical density of about 16.9g/cm$^{-3}$. This structure has dimensions consistent with a distorted A15 unit cell.[6] Later x-ray powder patterns were indexed by Miller[7] with a tetragonal unit cell, having $a$=5.34, $c$=9.92 Å which is in agreement with that found by Read and Altman. Das[8] interpreted the electron diffraction data on β-Ta as a bcc-based superlattice structure with slight tetragonal distortion. Nitrogen atoms located interstitially were an integral part of the unit cell, which has the dimensions of $a$=$b$=10.29, $c$=9.2 Å, and belongs to the *4/mmm* space group. On the other hand, Burbank[9] proposed a hexagonal subcell with $a$=2.831 and $c$=5.337 Å for β-Ta, which is related to a simple hcp structure with space group symmetry *P6$_3$/mmc*. Mosley *et. al.*[10] prepared β-Ta by electrodeposition and indexed the x-ray diffraction pattern in terms of a tetragonal unit cell with $a$=10.194, $c$=5.313 Å. The crystal structure of β-Ta was determined to be isomorphous with β-Uranium.

Our primary aim was therefore to investigate the structure of β-Ta from a local perspective in order to determine the most accurate structural model. All previous studies relied on x-ray or electron diffraction techniques. A structural probe which determines atomic structure on a short length scale can complement the long-range structural measurements previously performed.

In our study extended x-ray absorption fine structure (EXAFS) analysis was applied to probe the local structure of β-Ta. Theoretical calculations have been performed for all candidate models proposed for β-Ta. The polarized EXAFS and thermal effect analysis on both x-ray diffraction and EXAFS spectra were applied to evaluate the most likely candidate model determined by the comparison analysis of the EXAFS experimental data and theoretical calculations.

## II. EXPERIMENTAL AND COMPUTATIONAL METHODS

Beta tantalum coatings 25 μm thick were deposited on AISI4340 steel substrate by planar magnetron sputtering system in a 10 mTorr argon atmosphere at Benét Laboratories. The substrate temperature was ~350 ºC. Ta foils used as standards were obtained from Goodfellow Corporation and were 99.9% Ta. To determine the phase of the coatings, x-ray diffraction (XRD) was performed using a Philips x-ray diffractometer with Cu Kα radiation operated at 45 kV and 40 mA. XRD patterns for bcc tantalum powder (99.9%, Aldrich Chem. Co.) and foil were also collected for comparison under the same measurement conditions.

EXAFS measurements were carried out at beam line X-11A at the National Synchrotron Light Source, Brookhaven National Laboratory, using a Si(111) double crystal monochromator. For each measurement, the silicon double monochromator was detuned to approximately 80 percent of maximum transmitted x-ray intensity to reduce the harmonic content. The storage ring operated at 2.58 or 2.8 GeV with a typical current of 180 mA. Polarized x-ray absorption spectra were measured for the highly (002) textured beta samples. Three angles between the incident beam and sample surface were used, 90° (in-plane), 2.5° (grazing angle), and 45°. At low temperature (20 K), spectra were measured with the 45° incident beam for both β-Ta and α-Ta (Ta foil) and compared to data at 300 K. All the data were taken in fluorescence mode and collected at the L-III edge of Ta (9.732-10.986 keV). Energy calibration was accomplished by assigning the position of the first inflection point of the tantalum foil absorption spectrum to 9.881 keV. Nitrogen and argon gases were used respectively in the ion chamber and fluorescence detector to measure the incident ($I_0$) and fluorescence ($I_f$) intensities. For the α-Ta samples, six scans were collected, while for β-Ta 25 – 46 scans were measured for each orientation to improve statistics. The reference spectrum was obtained from six scans of α-Ta foil.

Theoretical fine structure $\chi(k)$ spectra for four proposed models were generated using FEFF7[11,12]. EXAFS data reduction and fitting were performed using WinXAS[13] following standard procedures[14]. Crystal structural drawings were generated using PowderCell[15] and CrystalMaker[16].

In the theoretical EXAFS calculations, the EXAFS amplitudes of the model spectra are larger than the corresponding experimental spectra. This significant difference is mainly due to the following two processes[14]: (a) inelastic losses such as multielectron excitation of the absorbing atom and the inelastic scatterings of the photoelectron and (b) static and thermal disorder effects. The inelastic and multielectron effects alter mainly the amplitude of the signal. The amplitude reduction factor $S_0^2$ was determined by measuring the α-Ta powder fluorescence and transmission data and then taking their ratio. To partially model the disorder effect we introduced a global Debye-Waller factor in *exp[-2$\sigma^2 k^2$]* term with $\sigma^2$=0.005 as obtained from fitting the experimental spectrum of bcc tantalum. The value of $\sigma^2$ was certainly underestimated because of the significant disorder in the β-Ta structure. Large disorders accompanied by asymmetric pair distributions or an anharmonic vibration often leads to a reduction of the EXAFS amplitude[14]. The *c(k)k³* functions were Fourier transformed (FT) into *r* space by using a Bessel window function for the same *k* range in each comparable data set.

### III. STRUCTURAL MODELS

Four structures proposed in the literature were considered (Table I): (A) a β-Uranium based structure; (B) a distorted A15 structure; (C) a bcc-Ta based superlattice structure with N interstitials; (D) simple hcp structure. The optimized Wyckoff-specific atomic positions were generated for all structure candidates and are shown in Fig.1.

Model A structure (Fig. 1(a), (b) ) is generated from the model of β-Uranium based on the work of Lawson[17] with space group *P4₂/mnm*. The lattice constants are from the index results of x-ray diffraction for β-Ta by Moseley[10]. The unit cell has five distinct Ta sites with a broad distribution of Ta-Ta distances about each site. The shortest Ta-Ta distance is 2.60 Å.

For the distorted A15 model [1,6,7], two structure types $B_1$ and $B_2$ were tentatively considered. In the $B_1$ structure, a cell with dimensions *a*=5.34, *c*=9.94 Å and general positions *a, c* and *e* were used. No other appropriate positions can be found in the International Tables which have the A15 symmetry with 16 atoms per unit cell. However, the shortest Ta-Ta distance for the constructed structure is 1.89 Å. This distance is nonphysical since tantalum metal has a radius of approximately 1.47 Å[18]. Another element in the transition metal group V, tungsten, exhibits a beta form with a simple cubic A15 unit cell containing eight atoms.[19, 20] Assuming that tantalum has a similar structure to β-tungsten, it may have a structure of the simple distorted A15 cell with *a*=5.34, *c*=4.97 Å, in which *c* is one half that found in the $B_1$ structure. The structure would have a density of 16.96 g/cm$^3$ and the shortest Ta-Ta distance is 2.67 Å. Here we regard this structure as one of the candidate structures called $B_2$ type (Fig. 1(c) ).

In model C (Fig. 1(d) ), β-tantalum was interpreted by the ordering of impurities in a bcc superlattice with elongation along the *a* and *b* directions.[8] Interstitial nitrogen was placed at octahedral sites at the centers of (010) and (100) faces. The Ta-N supperlattice has 56 atoms per unit cell, including two interstitial nitrogen atoms with a calculated density of 16.70 g/cm$^3$.

Model D (Fig. 1(e) ) is a hexagonal structure with ABAB stacking order similar to the structure of Zn or Cd.[9] The nearest neighbor Ta-Ta distance is 2.83 Å (within the atomic layer) and the next nearest Ta-Ta distance is 3.13 Å (between atomic layers). Table I summarizes the cell parameters, space groups, and other structural data for each model.

The details of α-Ta (bcc) and anomalous fcc structure are also listed for comparison[21]. The fcc phase of Ta was observed in very thin films of tantalum with the lattice constants of ~4.4 Å [22, 23]. This structure has a large shortest distance (3.1 Å), and low density (14.1 g/cm$^3$) as compared with other structures.

## IV. RESULTS AND DISCUSSION

### A. XRD analysis

The x-ray diffraction (XRD) patterns of the α-Ta (foil and powder samples) and deposited β-Ta film collected at room temperature are shown in Fig. 2. The intensities in β-Ta and Ta foil patterns were multiplied by a factor of 0.02 and 0.5, respectively. Compared to the Ta powder, the Ta foil shows a slightly preferred (110) orientation. For β-Ta film only three peaks which are referred to as the (002), (004) and (006) reflections were observed[10, 21]. The (002) peak is almost one hundred times greater than the (110) peak in the Ta powder pattern. The absence of additional peaks potentially suggests a high degree of texture exhibited by the β-Ta film. In fact, deposited β-tantalum thin films always exhibit (002) texture[24]. Pole figure analysis of β-Ta on glass reveal fiber texture normal to the film plane and confirm the (00*l*) assignment of the observed peaks[25]. Moreover, note that for bcc and β-U structures no (00*l*) reflections occur for odd *l*. The correspondence of this diffraction pattern with the local structure analysis will be discussed below.

**B. Comparative study of theoretical and experimental EXFAS spectra**

Detailed comparison of the experimental (room temperature) and theoretical EXAFS ($c(k)k^3$) and Fourier transform spectra are shown in Figs. 3 and 4, respectively. The amplitudes of the spectra were multiplied by appropriate factors for direct comparison. We compare the experimental (Fig. 3, curve (a) ) and theoretical (Fig. 3, curve (b) ) patterns for bcc Ta; they show the level of agreement anticipated. By comparing the patterns of candidate models (Figs. 3, curves (d) to (g) ) with experimental data (Fig. 3, curve (c) ) for β-Ta we can narrow the possible models for the structure of β-Ta. The model spectrum of β-U is consistent with the experimental spectrum of β-Ta structure for a broad range of $k$-space (to 10 Å$^{-1}$, at low $k$ values the Debye-Waller $s^2$ contributions to the amplitude are small). No other model matches the data over this broad region. We note that the fcc phase was not considered since it typically forms a thin layer (~50 Å) in the early growth phase [22, 23].

The Fourier transforms (FT) of the spectra (uncorrected for phase shift) are shown in Fig. 4. We can observe that among the four purposed models the spectrum based on β-U model (Fig. 4, curve (d) ) is the only one which shows a similar atomic short-range structure to the experimental data for β-Ta (Fig. 4, curve (c) ). In both spectra the major shell is located at ~ 2.9 Å with one sub-peak to the left. This shoulder in experimental data shows greater amplitude than that from theoretical calculation, but both are located at ~ 2.4 Å. All other models reveal large amplitudes at high $r$ resulting from highly ordered atomic arrangements as compared to the significant disorder (broad neighbor distributions) exhibited by the β-U model.

Efforts were taken to fit experimental data of β-Ta at room temperature (300 K) and low temperature (20 K) with a two shell model using eleven parameters: the coordination number $N$, atomic distance $R$, Debye-Waller factor $\sigma^2$, inner potential shift $\Delta E_0$ and third and fourth cumulants $C_3$, $C_4$. The amplitude reduction factor $S_0^2$ was fixed at 0.84 based on the ratio of α-Ta powder fluorescence and transmission data. However, fit results show that the cumulant expansion[26] does not converge up to the fourth order, which suggests high order terms or a more general distribution are needed to describe the atomic distribution in the first Ta shells. Significant local structural disorder exists, which cannot be modeled by a simple cumulant expansion. Hence, only the qualitative arguments above are given.

### C. Polarized absorption spectra of β-tantalum

Angular resolved EXAFS is a useful tool for detecting differences in the short-range structure in an anisotropic crystal. If β-Ta is isomorphous with β-Uranium, it would have a very different local structure along the $z$ direction compared to the $xy$ plane as shown in Fig. 1(a) and 1(b). For the (002) orientation of the β-Ta film, the spectrum acquired from a grazing angle (~2.5°) setup with the electric field perpendicular to the film approximately reveals local structure information along the $z$ direction of the β-Ta structure. The spectrum from the 90° angle (normal incidence) setup with the electric field parallel to the film provides information about the in-plane ($xy$) atomic distribution. The corresponding $k^3$ weighted $\chi(k)$ data are displayed in Fig. 5. Large differences exist between the spectra; the envelope amplitudes peak at different $k$ values. Fourier transforms of the two spectra are presented in Fig. 6 and are compared with their direct average and the conventional EXAFS data collected at 45°. The two polarized spectra depict quite different short-range atomic distributions. Dissimilar peak positions (nearest neighbor distances) and distinct amplitudes (coordination numbers) are exhibited for the first two or three shells. The high anisotropy is consistent with that expected of the β-U model. No other model yields asymmetry in the first shell. Differences between the averaged spectrum and that for the 45° measurement may be due to enhanced self absorption effects resulting in low amplitude in the grazing measurement.

### D. Analysis of thermal effects on XRD and EXAFS spectra

### (1) Thermal effects in the XRD pattern of β-Ta

As illustrated above, the most probable structure of β-Ta is tetragonal with the same space group of $P4_2/mnm$ as β-Uranium. To further investigate the structure, the calculated XRD based on β-U was performed as summarized in Table II. The values of the anisotropic thermal parameters obtained in space group $P4_2/mnm$ by Lawson[17] showed very little anisotropy. Therefore, in our case we simplified the temperature factor by using an isotropic form $exp(-2M)$,[27] where $M = 16\pi^2 \langle u_s^2 \rangle (Sin^2\theta)/\lambda^2$. The rms displacement $\langle u_s^2 \rangle$ is unknown for β-Ta, but the value for bcc-Ta at room temperature is approximately 0.015 Å$^2$, which was obtained from the powder diffraction data. Therefore, because β-Ta is more disordered than α-Ta, values of 0.04 and 0.01 Å$^2$ were used for β-Ta as a rough estimation. The values of $(1+cos^2 2\theta)/Sin\theta\, Sin2\theta$ are tabulated for Lorentz-polarization (LP) factors[27]. The calculation shows that the relative intensities with displacement $\langle u_s^2 \rangle^{1/2}$ at 0.2 Å, which is larger than that for in bcc-Ta, yet it is reasonable because of its highly disordered structure, agreeing well with experimental data.

The XRD pattern is an effective tool for assessing structure and has been extensively applied on β-Ta.[1-10, 21, 24] However, disagreements exist in the indexing results, which are mainly caused by differing relative intensities in diffraction patterns resulting from texture effects. Powder diffraction data with totally random orientation or single-crystal data are required for accurate assessment of the β-Ta structure.

**(2) Thermal effects in the EXAFS spectra of β-Ta**

EXAFS experiments at low temperature (20 K) and room temperature (300 K) were conducted for α-Ta and β-Ta in order to study thermal effects in the EXAFS spectra of β-Ta. The $c(k)k^3$ and FT spectra are displayed in Figs. 7 and 8 respectively. In Fig. 7, we observe that $c(k)k^3$ spectra at 20 K (dashed curve) have greater EXAFS signal than that at 300 K (solid curve). The signal enhancement is due to the decrease in thermal vibrations at low temperature. The amplitude change between 20 K and 300 K spectra in α-Ta is more significant than in β-Ta, indicating a large contribution of the thermal component in the Debye-Waller factor $\sigma^2$ for bcc-Ta.

Temperature effects on the local structure (Fig. 8) reveal α-Ta having large relative amplitude changes in both nearest shells (below 4 Å) and distant shells (above 4 Å) between 20 K (dashed curve) and 300 K (solid curve). Beta-Ta exhibits apparently smaller differences in amplitude at distant shells than these for the first few nearest shells between the two temperatures. This effect indicates that in β-Ta at low temperature the static disorder contribution, which is an intrinsic property of a system and temperature independent, is significant in high shells demonstrating the system's disorder. The disorder increases as *r* increases. The analysis shows that β-Ta has a highly disordered structure, which agrees with the low symmetry and anisotropic system of the β-U model.

## IV. CONCLUSIONS

Among the four proposed hypothetical structures, theoretically calculated EXAFS spectra based on the β-Uranium model are consistent with experimental data of β-Ta structure. The polarized EXAFS data verified the anisotropy of β-Ta, which is again in agreement with the β-U model. Temperature effects seen in XRD patterns and EXAFS spectra are consistent with the β-Uranium Model. Therefore, we conclude that β-Ta has a tetragonal structure with the most probable space group *P4$_2$/mnm*, containing 30 atoms per unit cell. Further experiments and quantitative analysis are needed to refine the exact atomic distribution of Ta atoms in this unit cell.

## ACKNOWLEDGMENTS

We thank the U.S. Army Sustainable Green Manufacturing Program for the financial support. Experimental data were collected at the National Synchrotron Light Source of Brookhaven National Laboratory funded by U.S. Department of Energy (DOE).

**TABLE I**. Lattice parameters of four hypothetical β-Ta structures, α-Ta and fcc-Ta.

| No. | Models | Space group | Lattice Constants (Å) | Shortest Ta-Ta distance (Å) | Density (g/cm$^3$) | Atoms / unit cell | Ref. |
|---|---|---|---|---|---|---|---|
| A | β-Uranium | $P4_2/mnm$ | a=10.194, c=5.313 | 2.60 | 16.33 | 30 | 10,17 |
| $B_1$ | Distorted A15 | $Pm(-3)n$ | a=5.34, c=9.94 | 1.89* | 16.96 | 16 | 1, 6, 7 |
| $B_2$ | | | a=5.34, c=4.97 | 2.67 | 16.96 | 8 | |
| C | Supperlattice Ta (N) | $P4/mmm$ | a=b=10.29, c=9.2 | 2.87 | 16.70 | 56 | 8 |
| D | Hexagonal | $P6_3/mmc$ | a=2.83, c=5.34 | 2.83 | 16.23 | 2 | 9 |
| α - Ta | | $Im(-3)m$ | a=3.31-3.33 | 2.86 | 16.55-16.27 | 2 | 21 |
| fcc-Ta | | $Fm(-3)m$ | a=4.39-4.48 | 3.10-3.17 | 14.21-13.37 | 4 | 21, 22, 23 |

* This Ta-Ta distance is nonphysical.

**TABLE II**. The calculation of relative intensities for (002), (004), (006) reflections in XRD base on β-U model.

| $2\theta$ | hkl | $\sin\theta/\lambda$ | $F^2(10^{-3})$ | LP Factor | Temp. Factor | | Rel. Intensity (Cal.) | | Rel. Intensity (Exp.) |
|---|---|---|---|---|---|---|---|---|---|
| | | | | | $\langle u_s^2 \rangle = 0.01$ | $\langle u_s^2 \rangle = 0.04$ | $\langle u_s^2 \rangle = 0.01$ | $\langle u_s^2 \rangle = 0.04$ | |
| 33.4 | 002 | 0.186 | 766.9 | 10.7 | 0.95 | 0.80 | 100 | 100 | 100 |
| 70.0 | 004 | 0.372 | 2230.2 | 2.07 | 0.80 | 0.43 | 47 | 30 | 28 |
| 118.7 | 006 | 0.558 | 318.5 | 1.63 | 0.61 | 0.14 | 4 | 1 | 1 |

## Captions

**FIG. 1**. Crystal structures of proposed models. The β-Uranium structure (Model A) from two orientations are shown in (a) and (b). The distorted A15 structure (Model $B_2$) is shown in (c). The bcc-based Ta-N supperlattice structure (Model C) is shown in (d) and the hexagonal hcp structure (Model D) is shown in (e).

**FIG. 2**. XRD patterns of Ta foil and β-Ta film. The diffraction of β-Ta film shows strong (002) reflection.

**FIG. 3.** *c(k) k³* spectra of α and β-Ta groups. (a) experimental (E) data for bcc(α)-Ta; (b) theoretical (T) data for bcc-Ta; (c) experimental data for β-Ta; and theoretical spectra of four hypothetical structure models: (d) β-Uranium model (Model A); (e) distorted A15 (Model $B_2$); (f) bcc-Ta based Ta-N supperlattice (Model C); (g) hexagonal structure (Model D). Note the similarities between the experimental spectrum and the theoretically calculated one for the bcc structure and between the experimental spectrum for β-Ta and the theoretically calculated one based on β-U model.

**FIG. 4**. Fourier transforms of *c(k) k³* spectra for α and β-Ta groups over the *k* range 2.86 to 14.32 Å$^{-1}$. (a) experimental (E) data for bcc(α)-Ta; (b) theoretical (T) data for bcc-Ta; (c) experimental data for β-Ta; and theoretical spectra of four hypothetical structure models: (d) β-Uranium model (Model A); (e) distorted A15 (Model $B_2$); (f) bcc-Ta based

Ta-N supperlattice (Model C); (g) hexagonal structure (Model D). Note the similarities between the experimental spectrum and theoretically calculated one for the bcc structure and between the experimental spectrum of β-Ta and theoretical calculated one based on β-U model.

**FIG. 5.** $c(k)\, k^3$ spectra of the polarized EXAFS experiment collected at angle of ~2.5° (grazing) and 90° (in-plane) for (002) orientated β-Ta film. Note the differences between the two spectra in terms of shape and peak positions of the amplitude envelope.

**FIG. 6.** Fourier transforms of $c(k)\, k^3$ spectra from the polarized EXAFS data (grazing and in-plane), the averaged spectrum and experimental data collected at 45° with $k$ range of 2.86 to 14.32 Å$^{-1}$.

**FIG. 7.** $c(k)\, k^3$ spectra of EXAFS experiment collected at temperature of 20 K (dashed curve) and 300 K (solid curve) for α-Ta and β-Ta film. Note the differences in the signal changes between two temperatures for both phases as $k$ increases.

**FIG. 8.** Fourier transforms (over the $k$ range of 2.81 to 15.78 Å$^{-1}$ for α-Ta and 2.86 to 14.32 Å$^{-1}$ for β-Ta) of $c(k)\, k^3$ spectra of EXAFS experiment collected at temperature of 20 K (dashed curve) and 300 K (solid curve) for α-Ta and β-Ta films.. Note the differences in the amplitude changes between two temperatures for both phases as $r$ increases. The large static disorder in β-Ta in high shells results in a small change in amplitude with temperature.

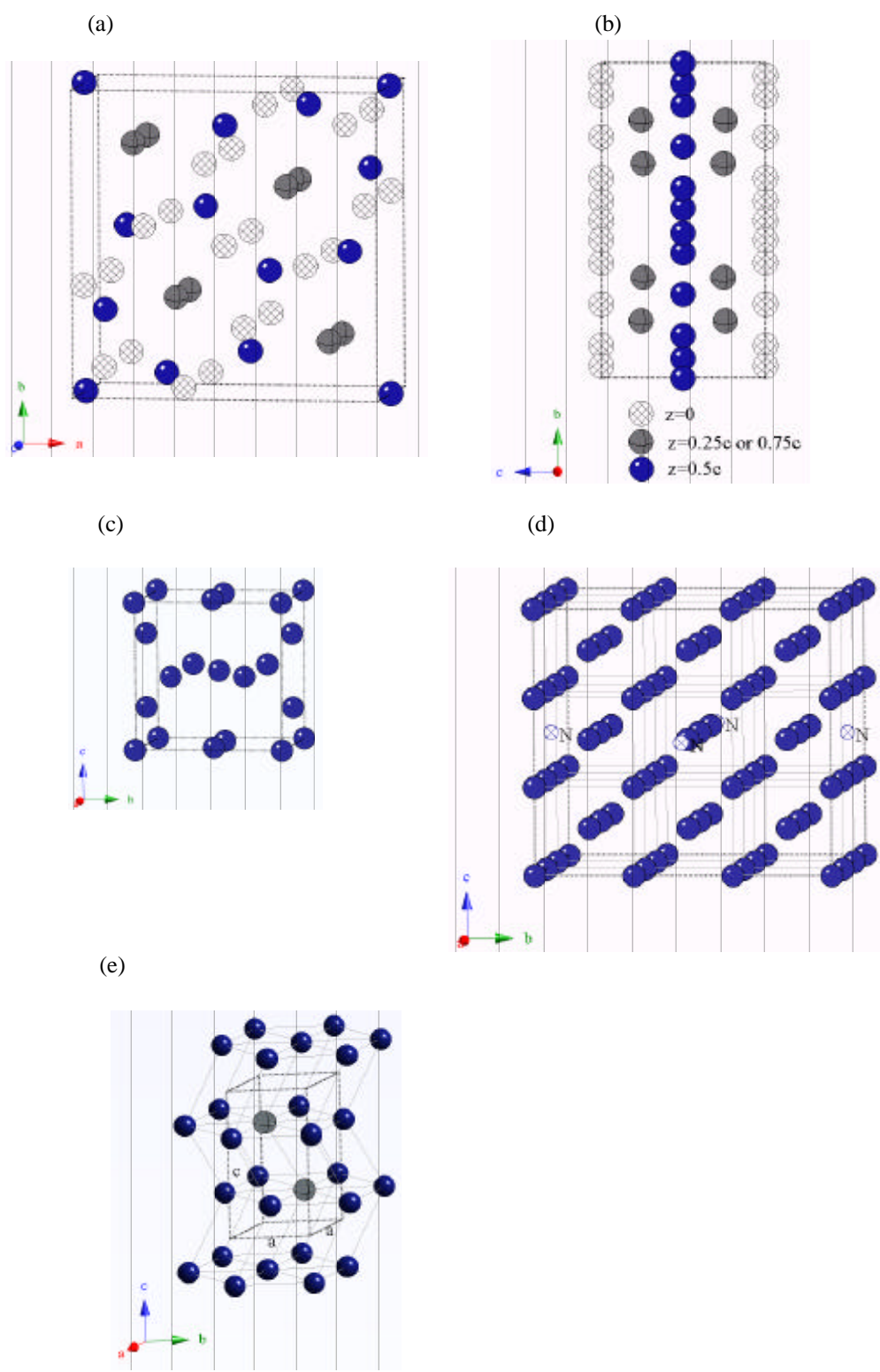

FIG. 1

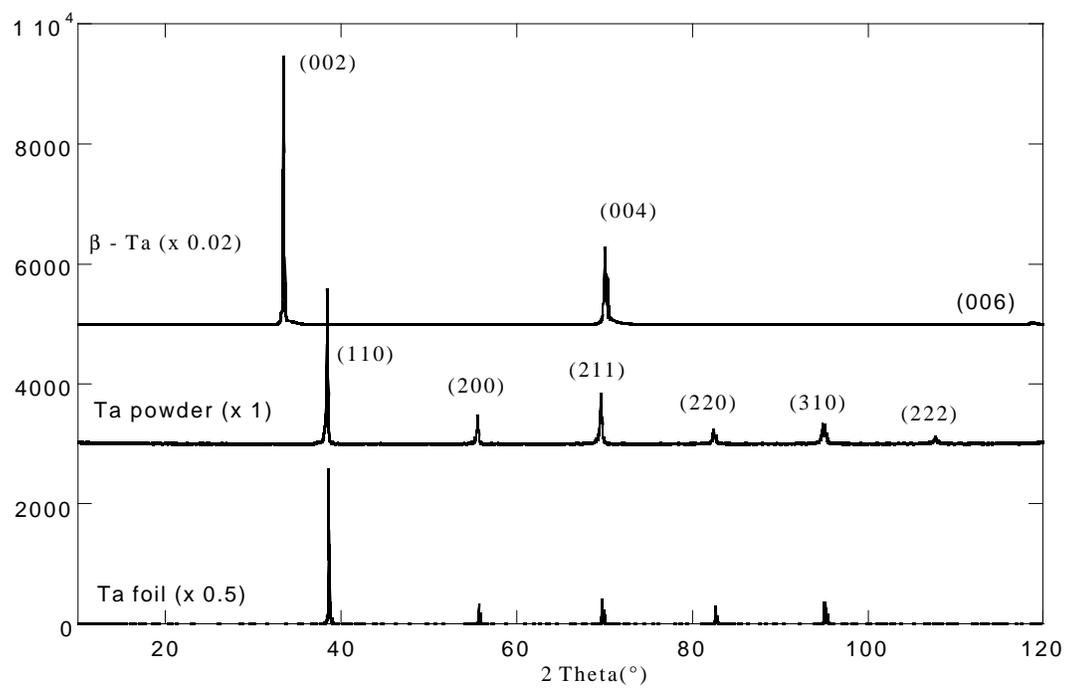

FIG. 2

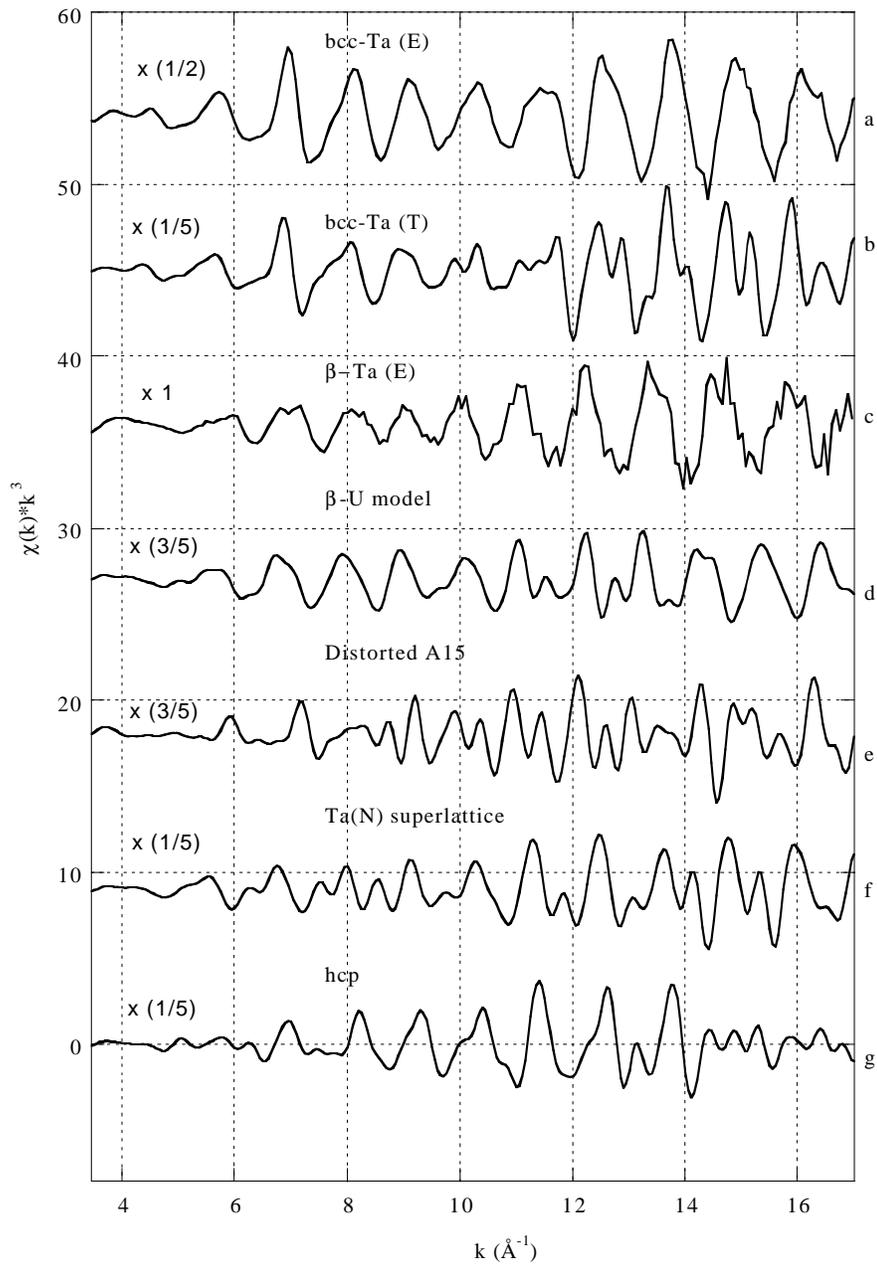

FIG. 3

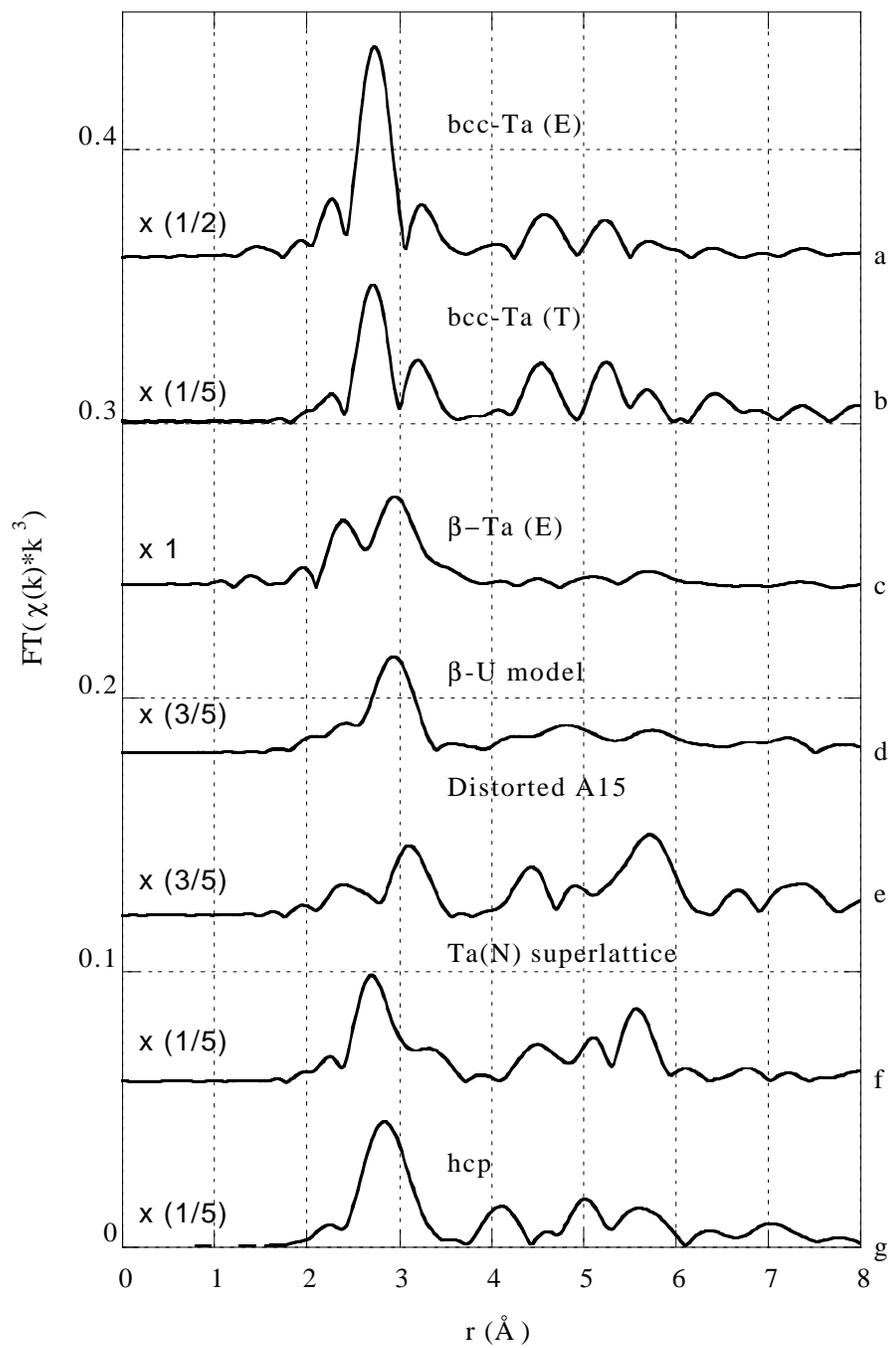

FIG. 4

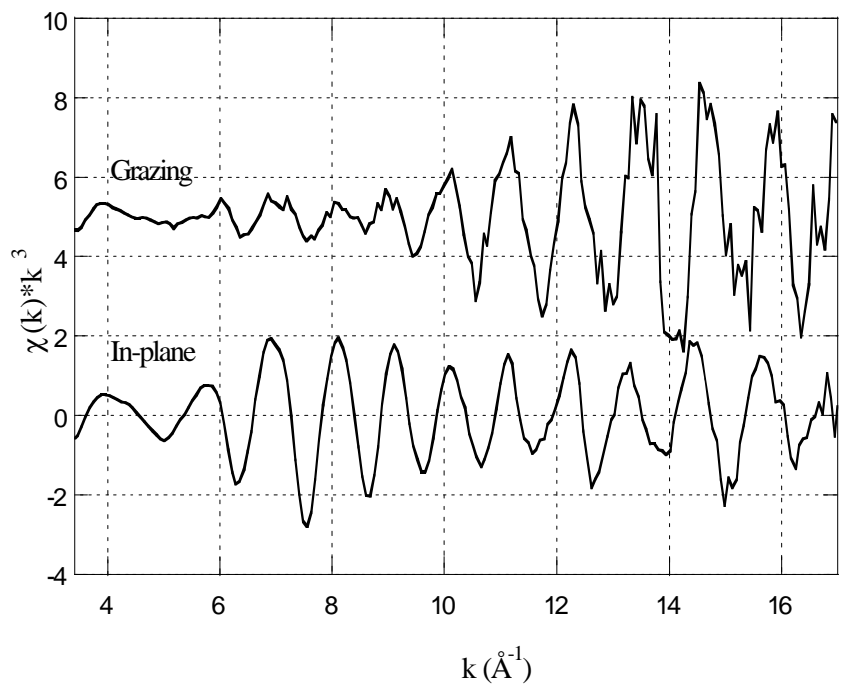

FIG. 5

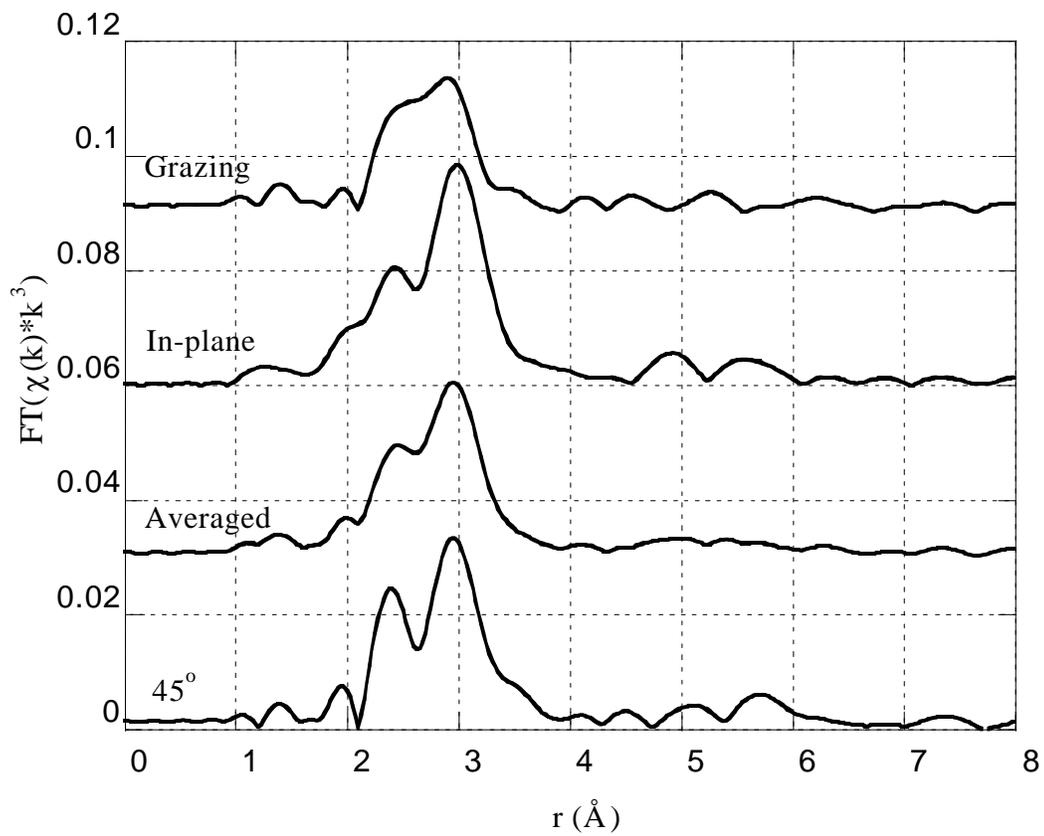

FIG. 6

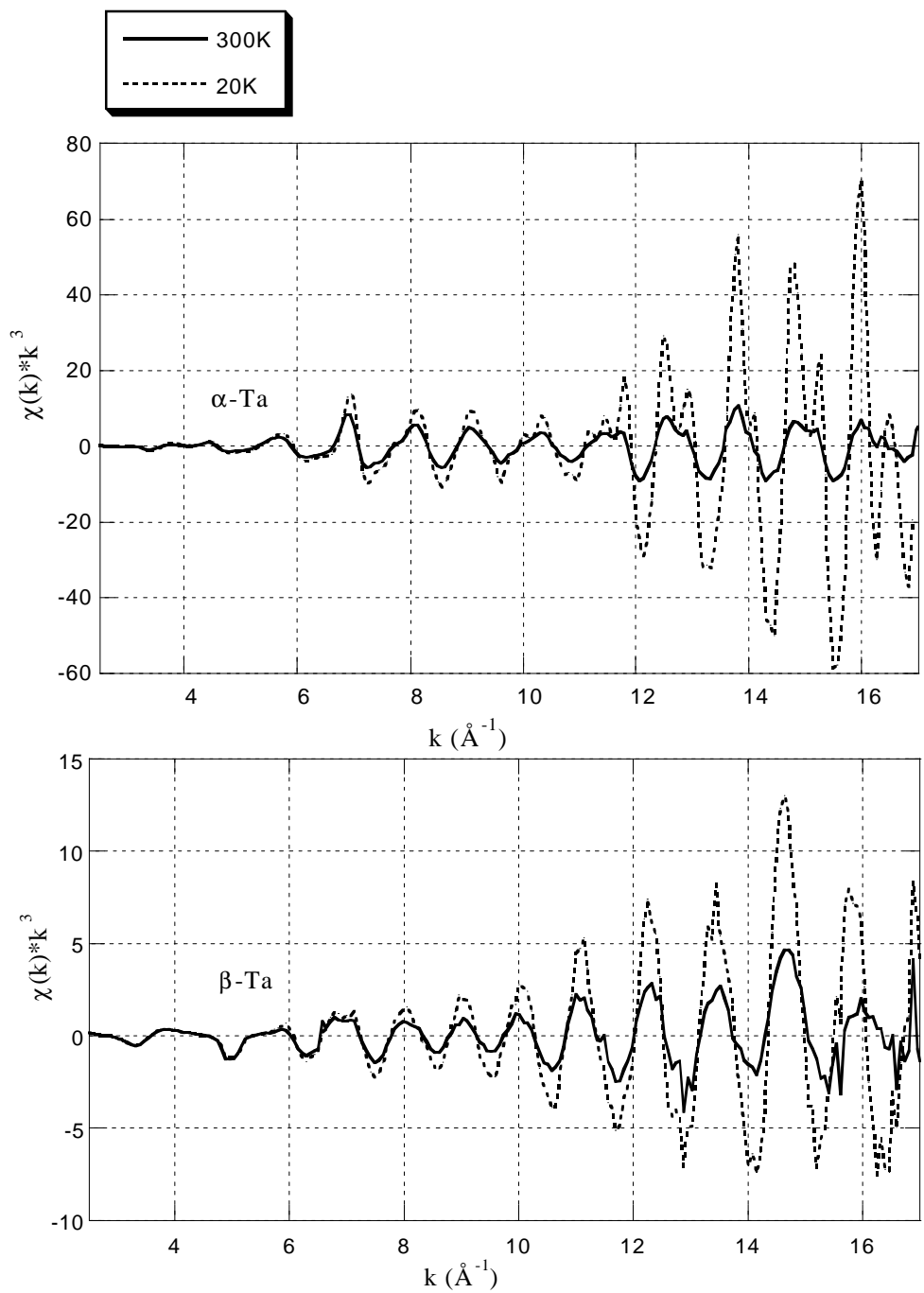

FIG. 7

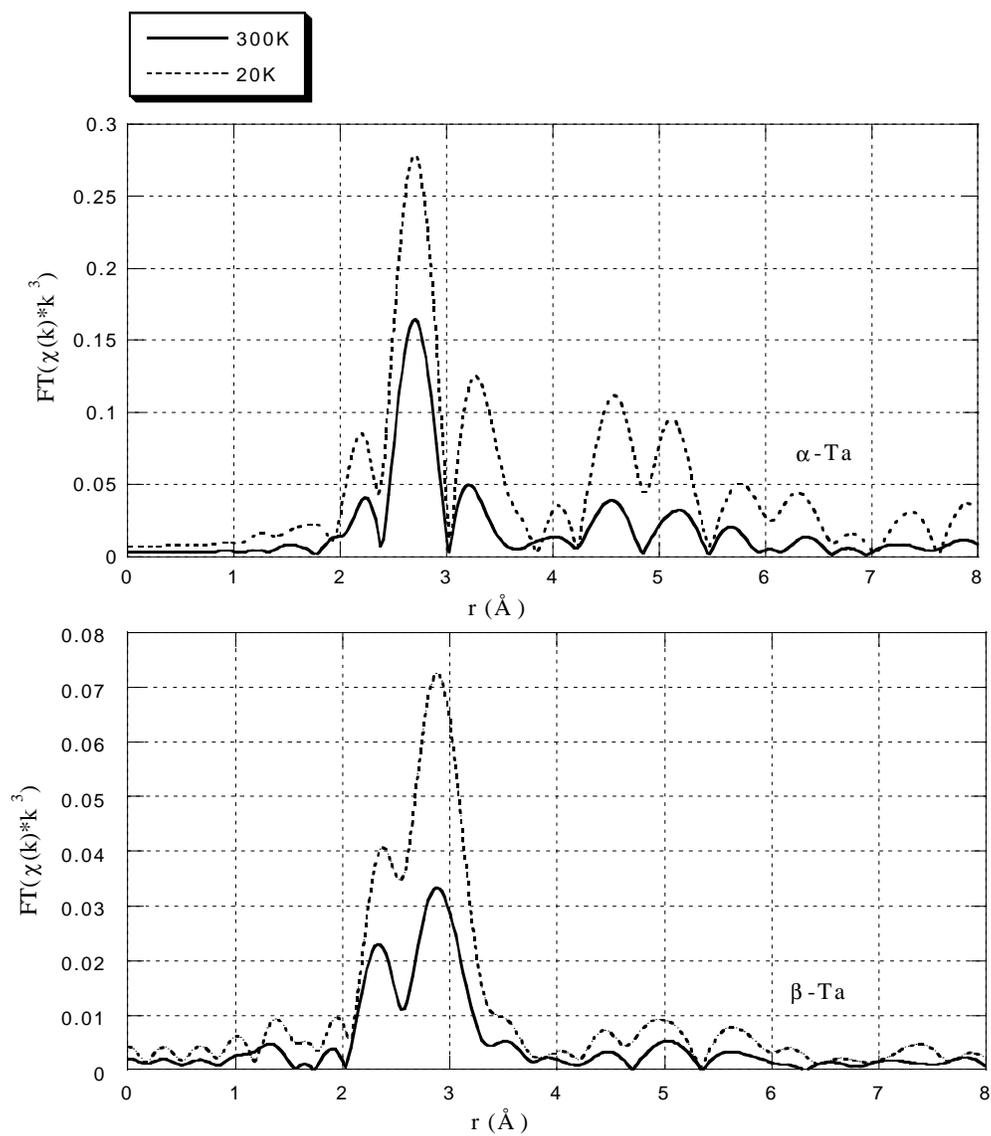

FIG. 8